\documentclass[10pt,twocolumn,conference]{IEEEtran}

\IEEEoverridecommandlockouts

\usepackage{graphicx}
\usepackage{amsmath}
\usepackage{amssymb}
\usepackage[caption=false]{subfig}
\usepackage[noadjust]{cite}
\usepackage{float}
\usepackage{algorithm}
\usepackage{bibentry}
\usepackage{balance}
\usepackage{algorithm}
\usepackage{algorithmicx}
\usepackage{algpseudocode}
\usepackage{xcolor}
\usepackage{subfig}
\usepackage{mathtools}

\usepackage{url}

\graphicspath{{img/}}

\begin{document}

\bstctlcite{IEEEexample:BSTcontrol}

\title{Height-Dependent Spectrum Activity Measurements and Modeling: A Case Study with FM Radio Bands\thanks{This research is supported in part by the NSF award CNS-2450593.}}

\author{Sung Joon Maeng$^*$, Amir Hossein Fahim Raouf$^\dagger$, Ozgur Ozdemir$^\dagger$, \.{I}smail G\"{u}ven\c{c}$^\dagger$, and Mihail L. Sichitiu$^\dagger$\\$^*$Department of Electrical and Electronic Engineering, Hanyang University, Ansan, South Korea\\
$^\dagger$Department of Electrical and Computer Engineering, North Carolina State University, Raleigh, NC\\
{sjmaeng@hanyang.ac.kr, amirh.fraouf@ieee.org, \{oozdemi,iguvenc,mlsichit\}@ncsu.edu}}

\maketitle

\begin{abstract}
The increasing demand for wireless connectivity necessitates advanced spectrum modeling to enable efficient spectrum sharing for next-generation aerial communications. While traditional models often overlook vertical variations in signal behavior, this paper proposes a height-dependent propagation model using a helikite-mounted software-defined radio (SDR). We collected extensive measurement data across the 88 MHz to 6 GHz range in both urban and rural environments. As a case study to validate our methodology, we focus on the FM radio band, which allows us to use publicly available transmitter locations and transmit power levels to facilitate comparisons between analytical with measurement results. We identify a clear transition from non-line-of-sight (NLoS) to line-of-sight (LoS) regimes at a specific altitude threshold and propose an altitude-dependent path loss model that incorporates this transition. Our results demonstrate that the proposed model significantly outperforms the standard free space path loss (FSPL) model in complex urban topologies, providing a more accurate framework for altitude-aware spectrum prediction and management across emerging aerial wireless technologies and bands.
\end{abstract}


\section{Introduction} \label{sec:intro}
As wireless cellular networks evolve to support advanced features such as high-speed data transmission, low-latency communication, and massive machine-type connectivity, the efficient utilization of available spectrum resources becomes highly critical. A significant challenge in this evolution is the increasing use of the vertical dimension by aerial platforms, which necessitates a deeper understanding of 3D signal behavior.
To manage these complex environments, concepts like Radio Dynamic Zones (RDZ) have been introduced to monitor and sense signals in real-time across specific geographical boundaries~\cite{maeng2022national}.
Realizing such concepts requires the ability to understand and predict signal patterns not just across time and 2D space, but also as a function of altitude.

Spectrum monitoring using USRPs has been widely investigated in the recent literature for ground-based applications~\cite{8254105,9395503,7343894}. In~\cite{8254105}, radar signal transmission and reception are conducted by utilizing two USRP N210 in the presence of Long-Term Evolution (LTE) and Wireless Local Area Network (WLAN) signals. In~\cite{9395503}, multiple standardized signals including LTE, radar, and WLAN are generated by USRPs and detected by using a machine learning approach. In \cite{7343894}, a USRP is employed to collect data and monitor the utilization of the spectrum in real-time, and a testing method has been proposed to measure the latency of the system. However, there is a lack of empirical modeling that specifically addresses the transition of propagation regimes for aerial receivers. While prior work has explored spectrum occupancy at various altitudes, many existing models fail to capture the precise height-dependent threshold where signal behavior shifts from shadowed to line-of-sight (LoS) conditions.


In this work, we deploy a helikite-mounted software-defined radio (SDR) to conduct a broad spectrum monitoring experiment from 88 MHz to 6 GHz in diverse urban and rural environments. To validate our proposed height-dependent propagation methodology, we utilize the frequency modulation (FM) radio band (87 MHz–108 MHz) as a strategic case study. This band is uniquely suited for validation because transmitter parameters—including precise locations, transmit power, and antenna patterns—are publicly accessible, providing a reliable ``ground truth" for analytical comparison that is often unavailable for proprietary 5G and 6G bands. The key contributions of this paper are as follows:
\begin{itemize}
\item Altitude-Dependent Path Loss Model: We identify a distinct altitude threshold (e.g., 50 m in urban areas) where signal propagation transitions from a Non-Line-of-Sight (NLoS) to an LoS regime.
\item Empirical Validation and Modeling: We propose a new altitude-dependent path loss model that incorporates this transition and demonstrate its superior accuracy over the standard free space path loss (FSPL) model.
\item Technology-Agnostic Framework: We demonstrate that our validation methodology using FM bands provides a scalable framework for spectrum prediction and management that can be extended to any emerging aerial technology and frequency band.
\end{itemize}

The rest of the paper is organized as follows. Section~II describes the experimental setup and measurement campaign. Section~III analyzes spectrum activity in the time and altitude domains. Section~IV presents the proposed altitude-dependent path loss model and compares it with measurement results, while Section~V concludes the paper.

\begin{figure}[!t]
        \centering
        \subfloat[Helikite floats over the experiment site.]{ \includegraphics[width=0.45\textwidth]{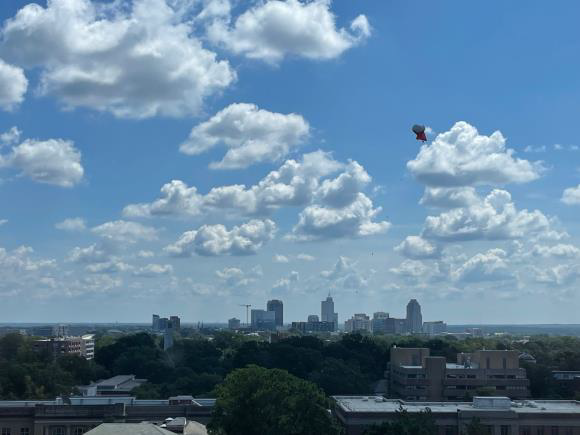}\label{fig:helikite}}
        
        \subfloat[Altitude of the helikite during the experiment in 2023 NC State Packapalooza festival.]{ \includegraphics[width=0.45\textwidth]{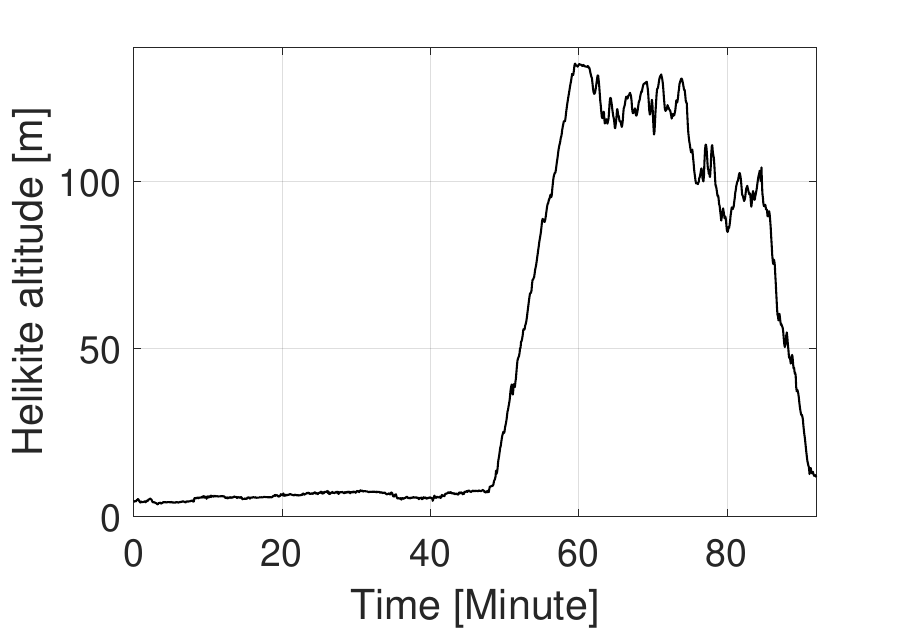}\label{fig:altitude}}\\
        \subfloat[PDF of the 2D distance from the 2D launch position of the helikite during the experiment.]{ \includegraphics[width=0.45\textwidth]{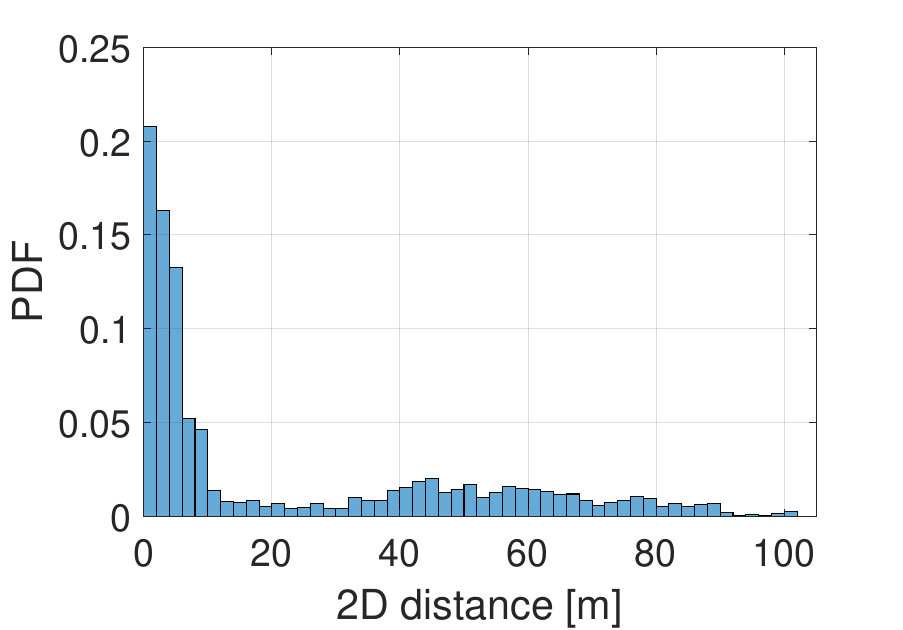}\label{fig:dist2D_pdf}}
	\caption{Photo of helikite flying over the experiment site, altitude and 2D distance change during the experiment.}\label{fig:heli_alt}\vspace{-3mm}
\end{figure}

\section{Spectrum Monitoring from Helikite} \label{sec:spec_moni}
We conduct a spectrum monitoring experiment using a helikite in an urban environment. We fly the helikite-mounted SDR and global positioning system (GPS) receivers around 90 minutes starting noon during NC State’s Packapalooza festival in August 2023. Results and analysis of the addtional experiments during NC State’s Packapalooza festival in August 2022 can be found in \cite{raouf2023cellular,maeng2023sdr,maeng2023spectrum}. Fig.~\ref{fig:helikite} shows the photo of helikite flying over the main campus of NC State University, Raleigh, NC.

\subsection{Measurement Campaign}
The helikite holds a low altitude of around 5~m for 45 minutes and then flies up to around 140~m. After the helikite holds its height for a few minutes, it gradually goes down to the ground. Although the spectrum is swept up to 6 GHz, we focus on the measurement of the FM radio frequency band from 87~MHz to 108~MHz in this paper.
A single spectrum sweeping takes around 15 seconds and a total of 364 sweeps are executed during the experiment. The altitude change of the helikite during the experiment processed by GPS logs is shown in Fig.~\ref{fig:altitude}. The helikite is tied by a tether to the ground and it floats around the experiment site due to the wind. Fig.~\ref{fig:dist2D_pdf} shows the probability density function (PDF) of the 2D distance from the initial position of the helikite during the experiment. 

In the rest of the paper, we will focus on the FM bands as a case study. However, our methodology can be extended to any band where transmitter locations, transmit powers, and antenna radiation patterns are known, e.g. cellular bands~\cite{CellMapper}. The reason for the specific focus on the FM spectrum is that, as highlighted in Table~\ref{table:FM_radi_sta}, information about signal sources can be easily obtained for this band. Since we rely on raw power measurements from all transmissions in the same band, we assume that interference from far away FM towers is negligible. To apply our proposed framework in other bands, received power from a specific signal source should be captured explicitly, e.g., through reference signal received power (RSRP) measurements for LTE and 5G transmissions.  

\begin{figure}[t]
	\centering\vspace{-4mm}
        \subfloat[Spectrum activity depending on the experiment time.]{\includegraphics[width=0.480\textwidth]{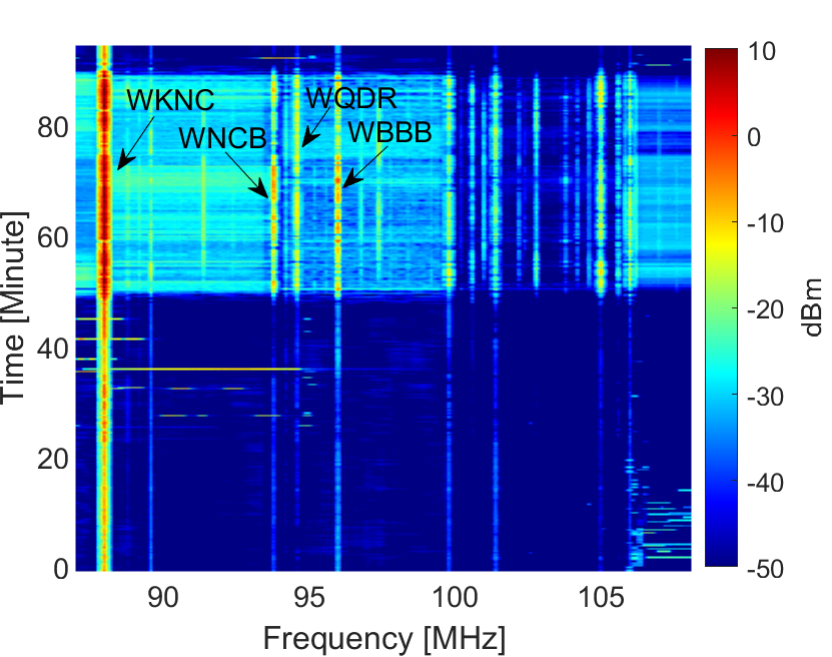}\label{fig:tim_spec}}
        
        \subfloat[Spectrum activity depending on the altitude of the helikite.]{\includegraphics[width=0.480\textwidth]{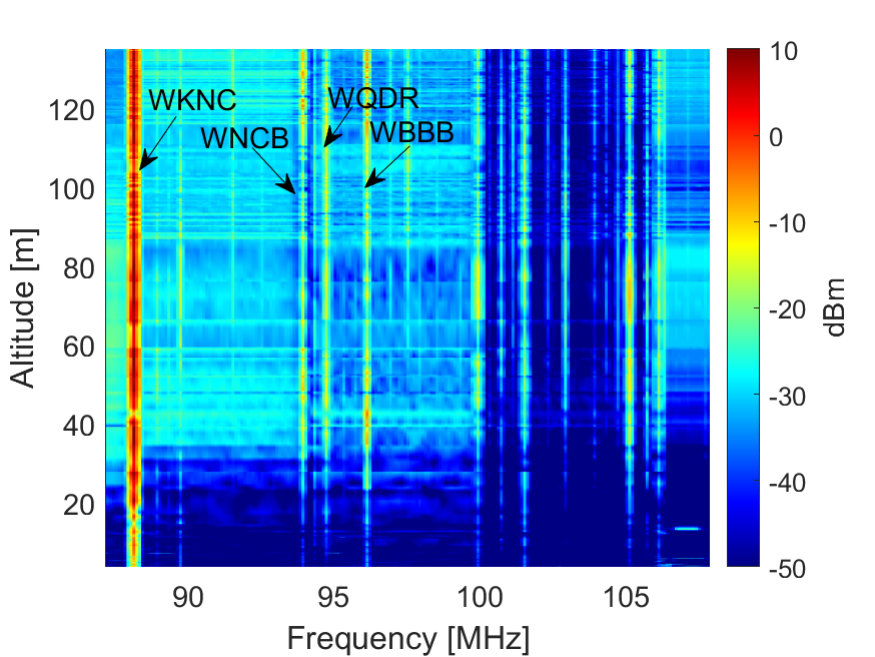}\label{fig:alt_spec}}
	\caption{Spectrum activity of the FM radio frequency band ($87$ - $108$~MHz) from the helikite measurement. Occupied spectrum for the WKNC, WNCB, WQDR, and WBBB FM radio stations are also highlighted.}\label{fig:spec_act}
\end{figure}
\begin{table}[!t]
\caption{List of four dominant FM radio stations near the experiment site~\cite{radio_locator}.}
\label{table:FM_radi_sta}
\centering
\begin{tabular}{p{0.7cm}|p{1.1cm}|p{0.9cm}|p{1.05cm}|p{1.3cm}|p{1.2cm}}
\hline
Call Sign & Frequency & FM tower height & Effective radiated power & Distance from Packapalooza & Distance from Lake Wheeler\\
\hline\hline
WKNC & $88.1$~MHz & \centering $59$~m & \centering $25000$~W  & $358$~m  & $7121$~m\\
\hline
WNCB & $93.9$~MHz & \centering $414$~m & \centering $100000$~W  & $15835$~m & $11052$~m\\
\hline
WQDR & $94.7$~MHz & \centering $507$~m & \centering $95000$~W  & $17026$~m & $15591$~m\\
\hline
WBBB & $96.1$~MHz & \centering $301$~m & \centering $98000$~W  & $12244$~m & $5137$~m\\
\hline\hline
\end{tabular}
\end{table}
\begin{figure}[!t]
	\centering
        \subfloat[3D map view of WKNC obtained by Google Earth.]{\includegraphics[width=0.47\textwidth]{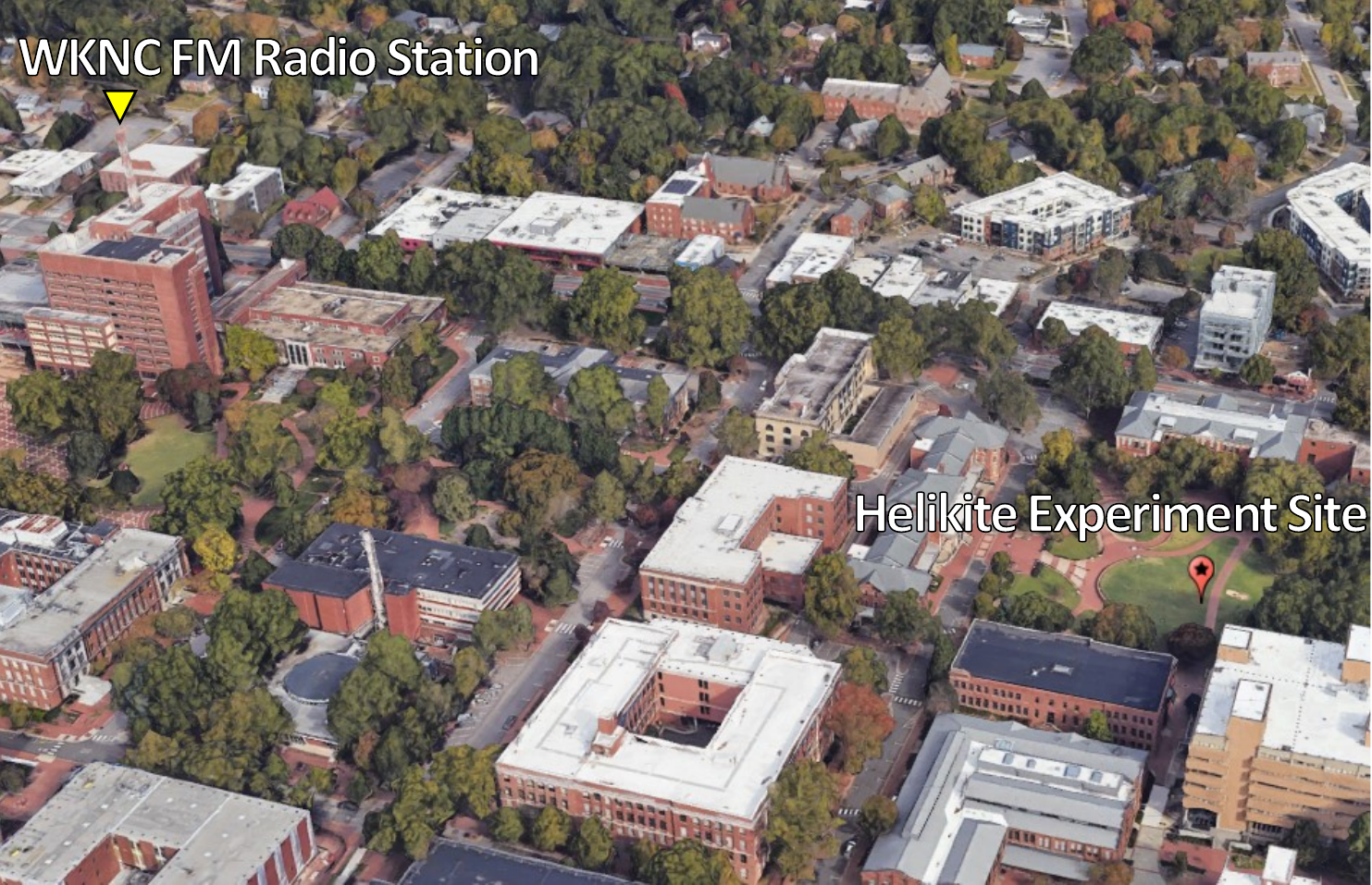}\label{fig:wknc}}

        \subfloat[Contour maps of WKNC.]{\includegraphics[width=0.47\textwidth]{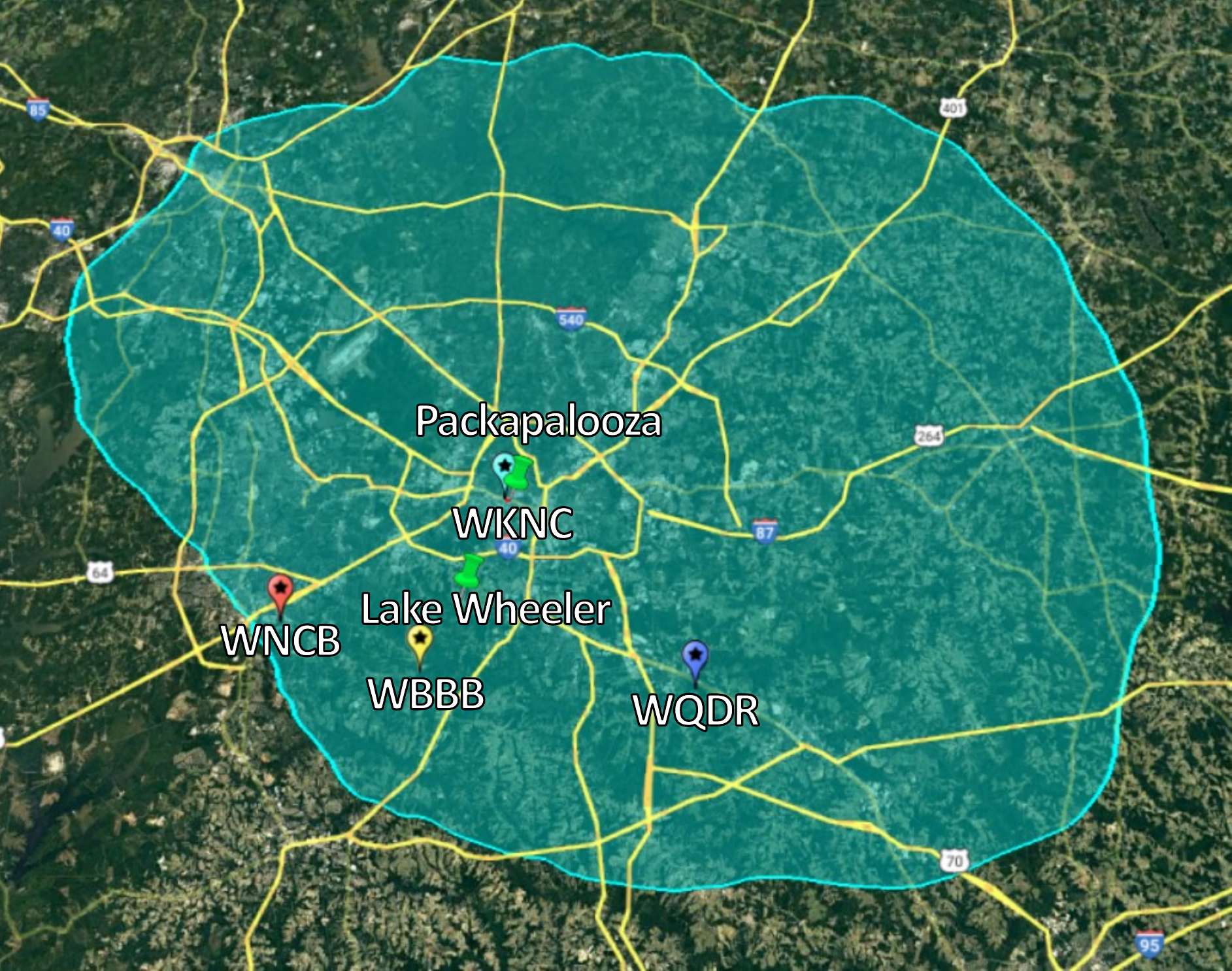}\label{fig:wknc_cov}}

        \subfloat[Contour maps of WNCB (yellow), WQDR (blue), and WBBB (red).]{\includegraphics[width=0.47\textwidth]{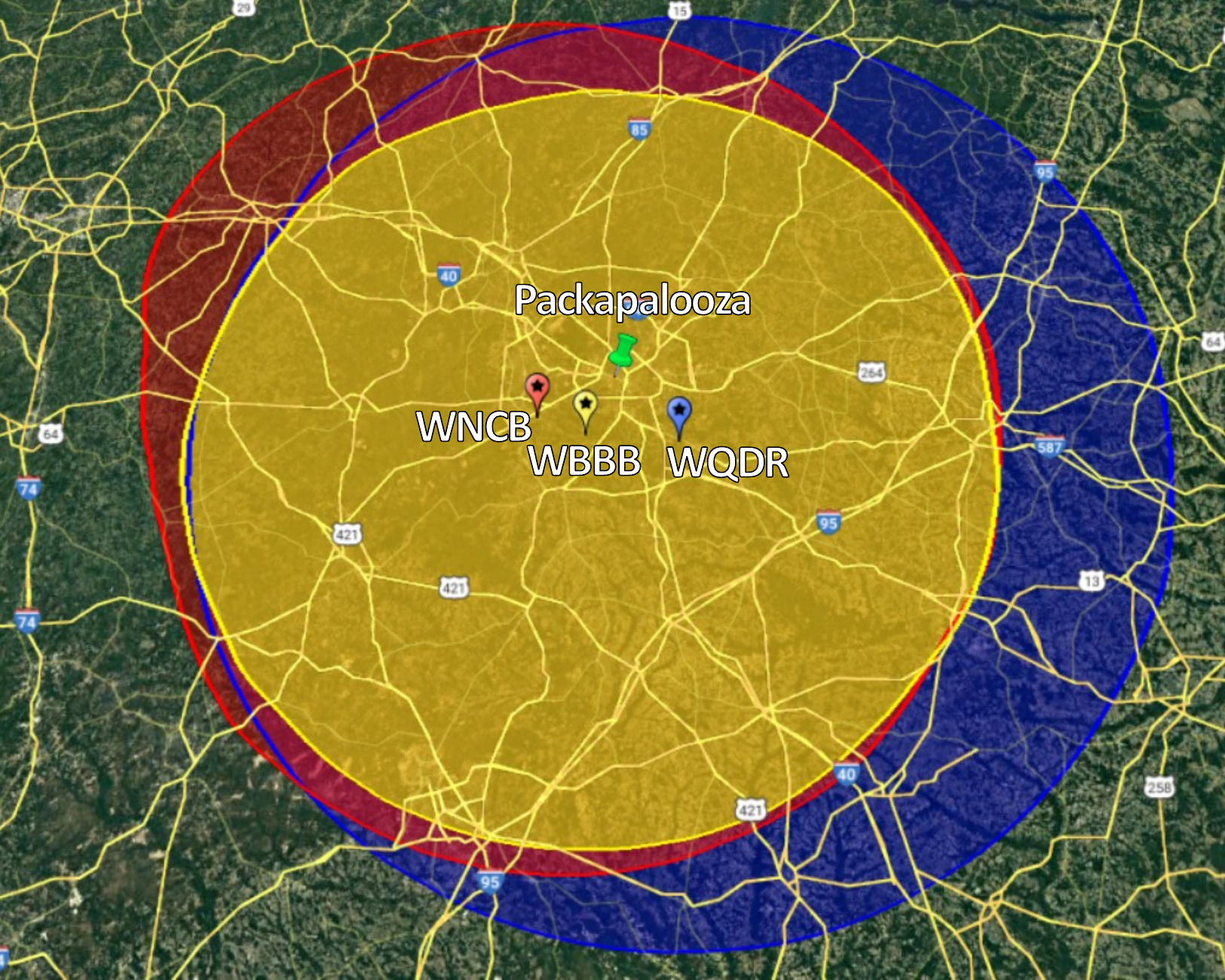}\label{fig:others_cov}}
	\caption{Location and coverage of the FM radio stations and the helikite experiment site location at the NC State Main campus.}\vspace{-5mm}
\end{figure}

\section{Spectrum Activity in FM Radio Band}\label{sec:spec_acti}
In this section, we present spectrum measurements for the FM radio frequency band ($87$~MHz - $108$~MHz) obtained using the heilkite platform. We post-process the spectrum measurement dataset from the SDR receiver and the altitude log from the GPS receiver using MATLAB.

\subsection{Spectrum Activity in Time and Altitude Domain}
Fig.~\ref{fig:tim_spec} and Fig.~\ref{fig:alt_spec} show the power spectrum of FM radio frequency band versus time (in minutes) and altitude (in meters). Throughout the paper, we present our results after compensating for the calibration error of the SDR receiver by adding $-34$~dBW on top of the measurement power based on the path loss analysis in Table~\ref{table:fre_spac} in Section~\ref{sec:fre_spa}. We observe that received signal strength is relatively high during the time period from 50 to 90~minutes when the helikite flies at higher altitudes. In addition, as the altitude of the helikite increases, received signal power gradually increases at low altitudes and then becomes constant from a certain altitude. Moreover, we observe four dominantly occupied signal sources in the monitored band, which come from FM radio stations: WKNC ($88.1$~MHz), WNCB ($93.9$~MHz), WQDR ($94.7$~MHz), and WBBB ($96.1$~MHz).  Public information about the radio stations used in spectrum analysis is summarized in  Table~\ref{table:FM_radi_sta}.

In particular, WKNC, the strongest signal observed, is operated by NC State University and is relatively close to the experiment site.  It implies that signal strength from the FM radio station is relatively strong at the helikite. Fig.~\ref{fig:wknc} shows the 3D sky view of the area where the locations of the helikite (Packapalooza festival) and WKNC FM radio station are indicated. Fig.~\ref{fig:wknc_cov} shows the contour maps of the coverage area of WKNC signal provided in~\cite{FCC_public_file}, which highlights the directional antenna pattern of the radio station. Fig.~\ref{fig:others_cov} shows the overlap contour maps of the coverage area of WNCB, WQDR, and WBBB, which are relatively broader due to their higher transmit power and are omni-directional. 

\vspace{-3mm}

\section{Proposed Heigh-Dependent Propagation Model}\label{sec:ana_radi_prop}
In this section, we analyze the altitude-dependent LoS/NLoS conditions and the received signal strength based on the free space path loss model, and introduce a new analytical model to capture altitude dependence. 

\subsection{ Analysis of Altitude-Dependent Path Loss Model}
\begin{figure}[t!]
	\centering
        \subfloat[2023 Packapalooza (urban) measurement.]{\includegraphics[width=0.5\textwidth]{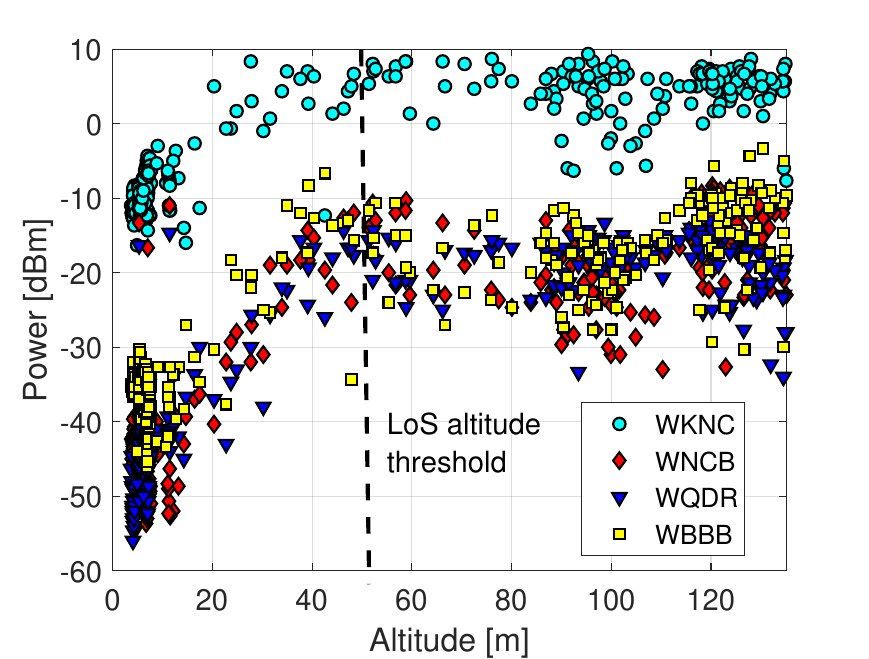}\label{fig:los_alt_Pa}}

        \subfloat[2023 Packapalooza (urban) analytical results using the proposed model in \eqref{eq:PL_bp}.]{\includegraphics[width=0.5\textwidth]{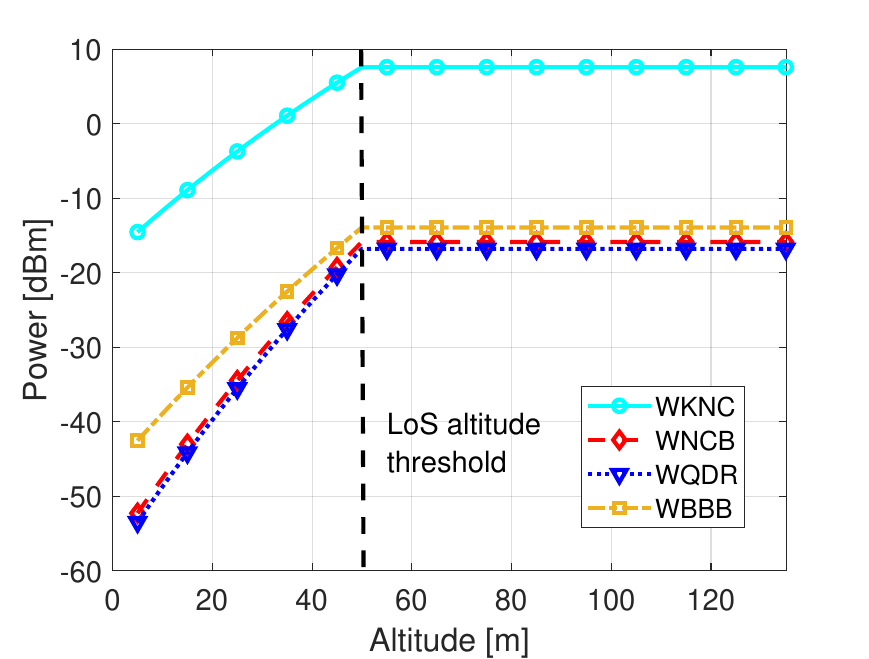}\label{fig:los_alt_Pa_ana}}

        \subfloat[CDF of the modeling error from 2023 Packapalooza spectrum data (urban) using \eqref{eq:PL_bp}, \eqref{eq:PL_fs}.]{\includegraphics[width=0.5\textwidth]{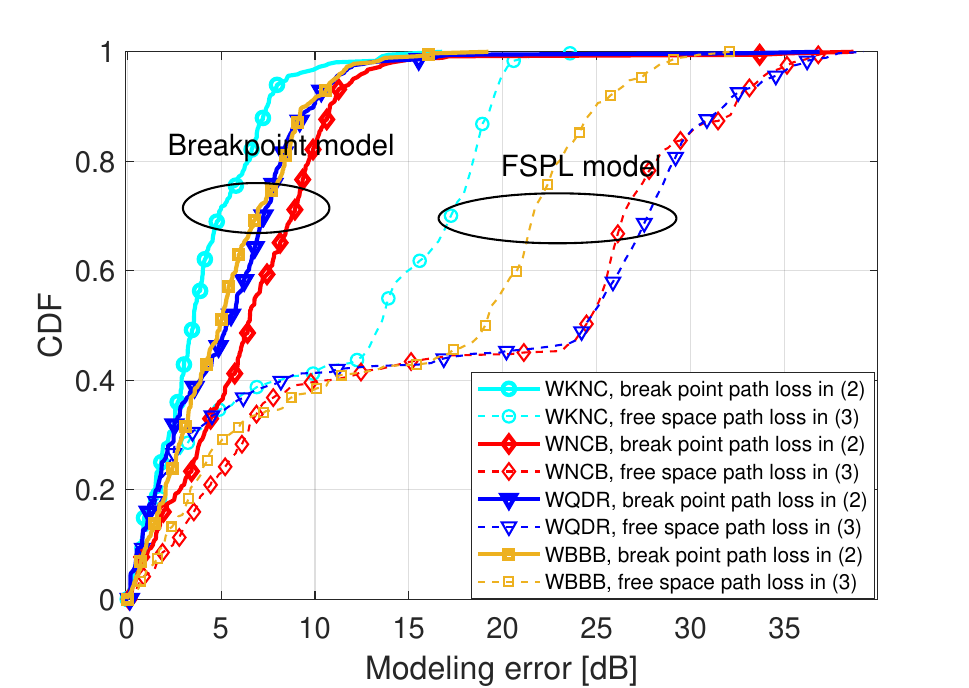}\label{fig:cdf_err_packa}}
	\caption{Power of the signals from different FM radio stations and corresponding altitude is marked for every spectrum sweep in the urban area. We obtain the LoS altitude threshold of the helikite as $50$~m based on the trend of the measurement points in the urban area.}\vspace{-5mm}
\end{figure}

\begin{figure}[t!]
	\centering
        \subfloat[2022 Lake Wheeler (rural) measurement.]{\includegraphics[width=0.5\textwidth]{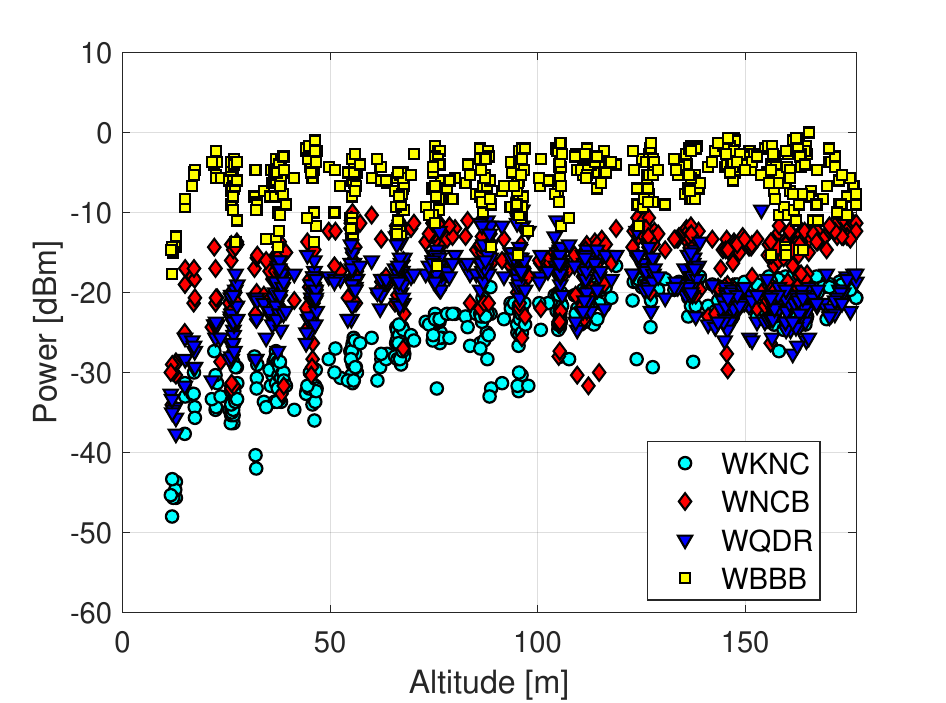}\label{fig:los_alt_LW}}

        \subfloat[2022 Lake Wheeler (rural) analytical results using the proposed model in \eqref{eq:PL_bp}, \eqref{eq:PL_fs}.]{\includegraphics[width=0.5\textwidth]{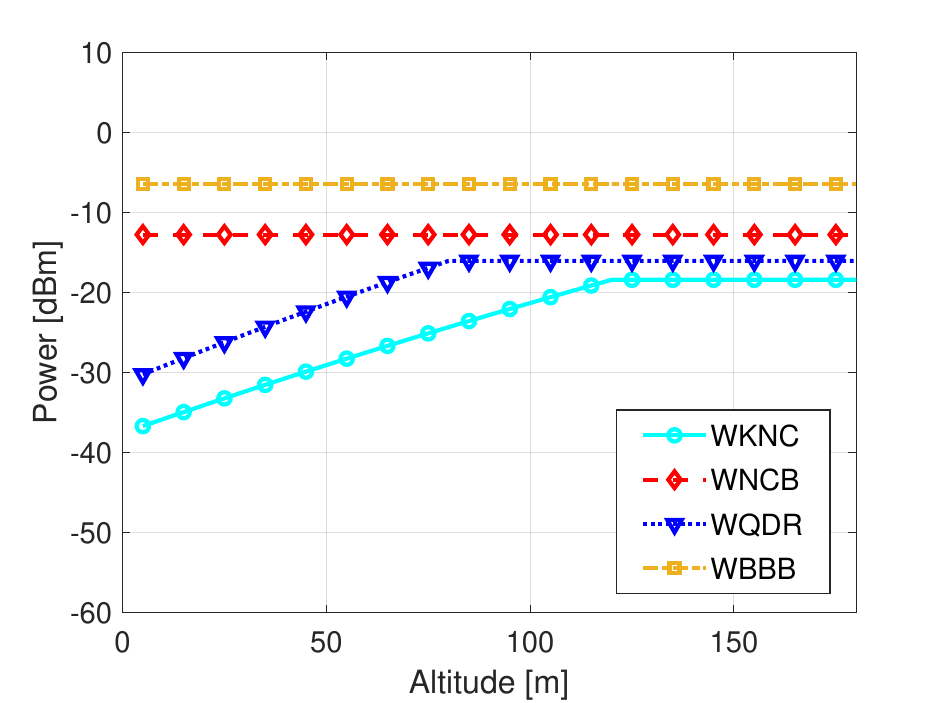}\label{fig:los_alt_LW_ana}}

        \subfloat[CDF of the modeling error from 2022 Lake Wheeler spectrum data (rural) using \eqref{eq:PL_bp}, \eqref{eq:PL_fs}.]{\includegraphics[width=0.5\textwidth]{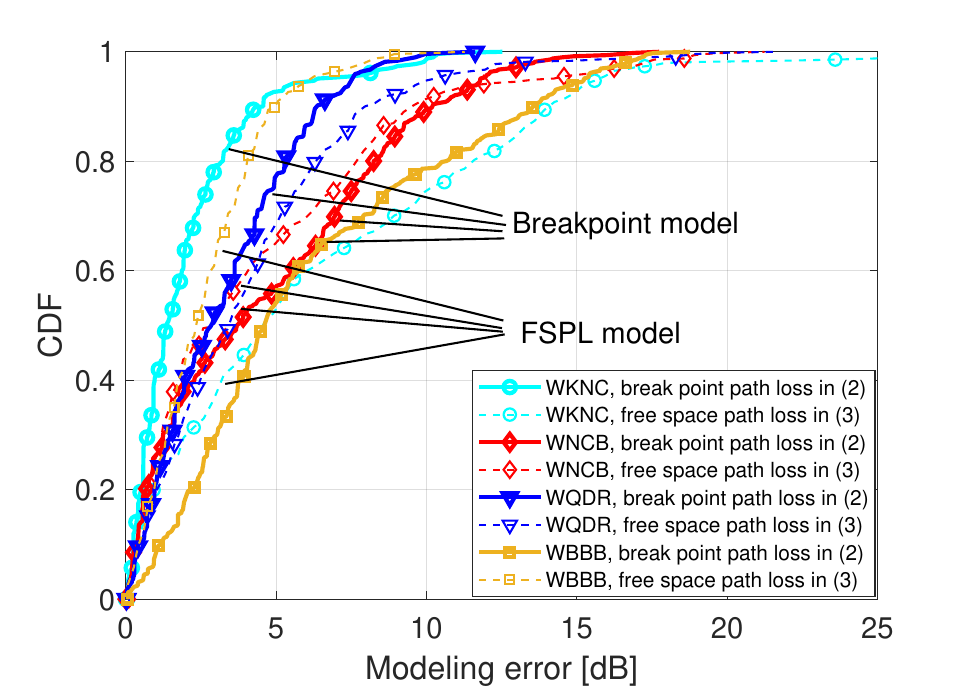}\label{fig:cdf_err_lw}}
	\caption{Power of the signals from different FM radio stations and corresponding altitude is marked for every spectrum sweep in the rural area. The path loss model fits either the free space path loss model or the altitude-dependent path loss model based on the location of FM radio stations. }\vspace{-3mm}
\end{figure}

\begin{figure}[!t]
	\centering
        \subfloat[Mean and standard deviation of power spectrum on LoS condition where the frequency of FM radio stations are indicated.]{\includegraphics[width=0.5\textwidth]{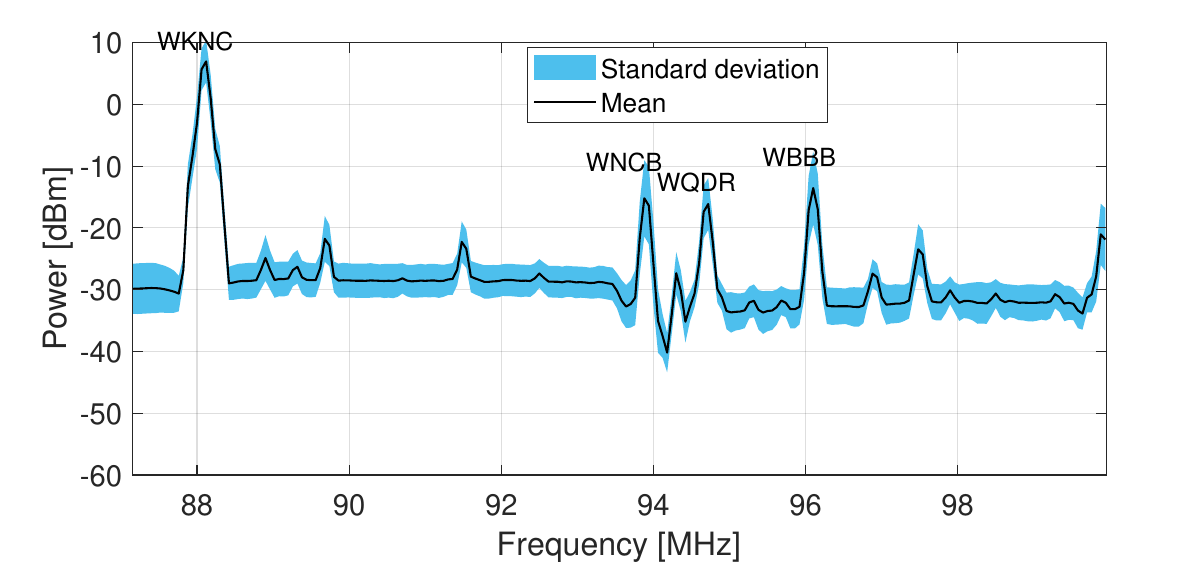}\label{fig:spec_mean_los}}
     
        \subfloat[CDF and median of power spectrum on LoS condition.]{\includegraphics[width=0.5\textwidth]{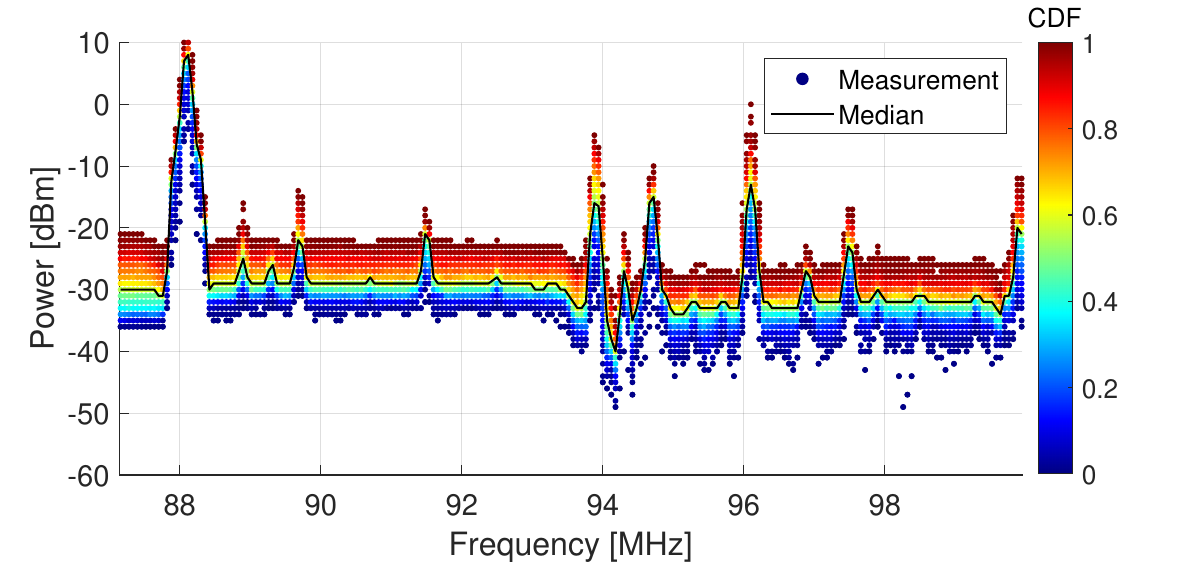}\label{fig:spec_cdf_los}}
  
        \subfloat[Mean and standard deviation of power spectrum on NLoS condition.]{\includegraphics[width=0.5\textwidth]{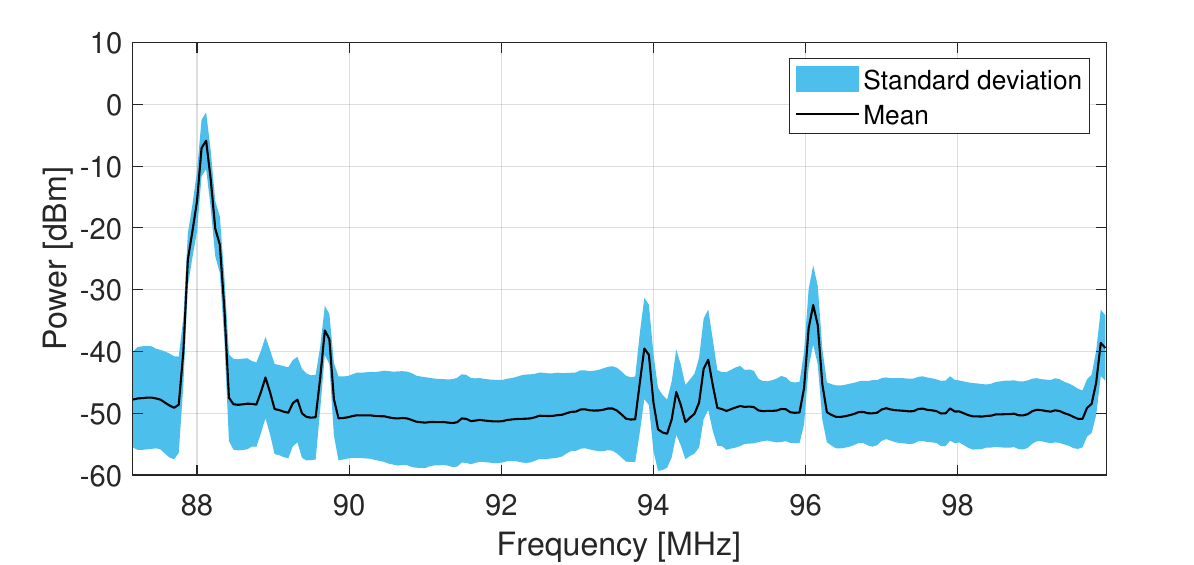}\label{fig:spec_mean_nlos}}
     
        \subfloat[CDF and median of power spectrum on NLoS condition.]{\includegraphics[width=0.5\textwidth]{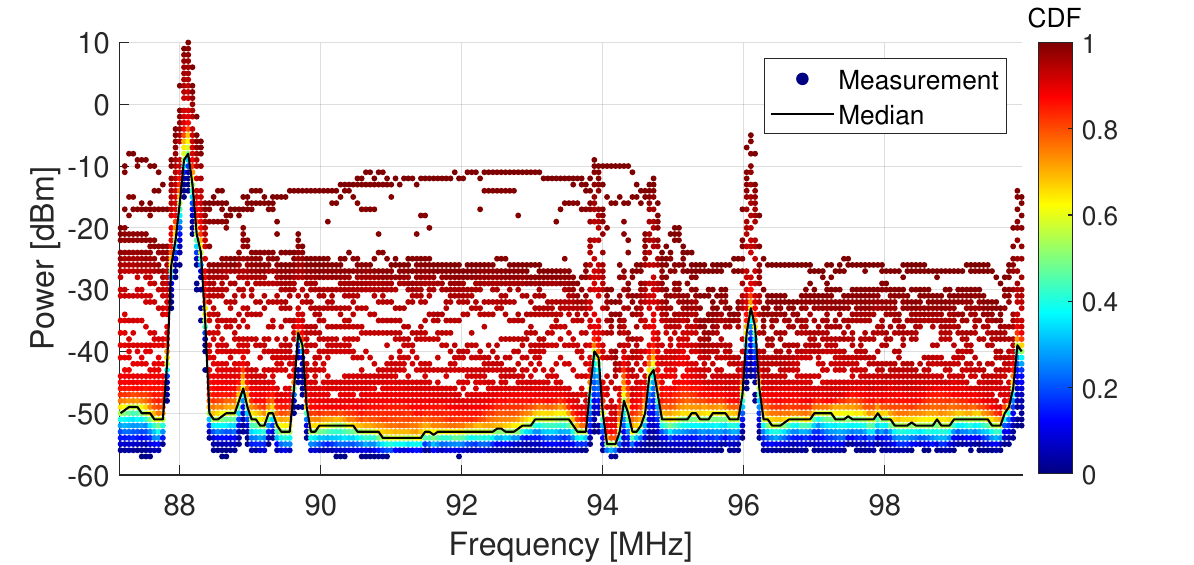}\label{fig:spec_cdf_nlos}}
	\caption{Spectrum occupancy of the FM radio band depending on the LoS/NLoS conditions represented by mean, standard deviation, median, and CDF function. The LoS (NLoS) condition considers measurements above (below) 50~m height.}\label{fig:spec_los_nlos}\vspace{-3mm}
\end{figure}

In radio propagation, the LoS/NLoS condition is decided by the blockage of the obstacles between the link. When the altitude of the helikite is low, the signals from the FM radio stations are blocked by buildings or trees and the link becomes NLoS. However, as the helikite flies up and the altitude of it becomes higher than the obstacles, the link becomes LoS. Therefore, as the altitude of the helikite increases from the ground level, the received signal strength increases until it reaches the LoS condition. After that altitude, the received signal strength becomes relatively constant. Note that since the horizontal distances between the radio stations and the helikite are far longer than the helikite altitude, an increment of the 3D distance between signal sources and the helikite by increasing altitude would be relatively small, which implies that pathloss increase by the altitude becomes relatively small. 

Based on the above observations, we propose an altitude-dependent break point path loss model based on NLoS and LoS conditions. We also analyze how well the proposed model fits to the measurement. In particular, the proposed model is given by
\begin{equation}
\mathsf{PL}(h)= \left\{ \,
\begin{IEEEeqnarraybox}[][c]{l?s}
\IEEEstrut
\mathsf{PL}_0 + 10\left(2e^{\frac{1}{\alpha}}e^{-\frac{h}{h_{0}\alpha}}\right)\log_{10}\frac{d}{d_0},&$h<h_0$ \\
\mathsf{PL}_{\rm fs}, & $h\geq h_0$
\IEEEstrut
\end{IEEEeqnarraybox}
\right.,\label{eq:PL_bp}
\end{equation}
where $h$, $h_0$, $d$, $d_0$, $\alpha$ denote the altitude,  the LoS altitude threshold, the distance,  the reference distance ($d_0=1$~m),  and the tunable fitting parameter, respectively. This formulation ensures a smooth and continuous transition between propagation regimes, avoiding artificial discontinuities while preserving the underlying physical trends observed in measurements. In addition, $\mathsf{PL}_{\rm fs}$ and $\mathsf{PL}_{0}$ indicate free space path loss at distance $d$ and free space path loss~(FSPL) at reference distance $d_0$, respectively. The former is given as
\begin{align}\label{eq:PL_fs}
    &\mathsf{PL}_{\rm fs}=\mathsf{G_{Tx}}+\mathsf{G_{Rx}}+10\log_{10}\left(\frac{4\pi d}{\lambda}\right)^2,
\end{align}
where $\mathsf{G_{Tx}}$, $\mathsf{G_{Rx}}$, $\lambda$ denotes the transmitter (Tx) antenna gain, the receiver (Rx) antenna gain, and the wavelength. In \eqref{eq:PL_bp}, the pathloss exponent gradually decreases to 2 until the altitude threshold $h_0$, after that altitude, the pathloss becomes FSPL.

\section{Numerical Results}\label{sec:num_res}
In this section, we analyze the radio propagation of signals from FM radio stations using helikite spectrum measurement results. We apply the proposed altitude-dependent path loss model to helikite measurements conducted in both urban and rural environments. In addition, we analyze the spectrum activity of the FM radio band under LoS and NLoS conditions. Finally, we obtain the offsets caused by calibration errors based on free-space path loss analysis.

\subsection{Path Loss Analysis on Urban Measurement Datasets}
Fig.~\ref{fig:los_alt_Pa} shows the received power levels from several FM radio stations as a function of helikite altitude. Each marker represents a measurement point from one of 364 spectrum sweeps. A clear pattern emerges: received power increases with altitude up to approximately 50~m, after which it plateaus. This trend reflects a transition from a NLoS regime where signal blockage from buildings and other urban structures is prevalent to a LoS regime above the rooftop level.
Based on this trend, we define an empirical LoS altitude threshold at 50~m. Only the data points above this threshold are used to fit the proposed path loss model, ensuring the modeling assumptions align with LoS propagation conditions.

Results in Fig.~\ref{fig:los_alt_Pa} highlight a key insight that is often overlooked in prior work. Altitude plays a critical role in shaping signal propagation, even for relatively low-frequency signals such as FM radio. By explicitly identifying and modeling the LoS transition point, our analysis addresses a gap in conventional path loss modeling, which typically assumes fixed-distance-based regimes and overlooks vertical variation in signal behavior.
The modeling results using the break-point path loss model in~\eqref{eq:PL_bp} are shown in Fig.~\ref{fig:los_alt_Pa_ana}. The fitting parameter $\alpha$ is set to $2.5$ for WKNC, WNCB, and WQDR, and to $3$ for WBBB. More precisely, we apply a nonlinear least squares curve fitting approach using the \texttt{lsqcurvefit} function in MATLAB to estimate the parameters of the proposed altitude-dependent path loss model in \eqref{eq:PL_bp}. This fitting procedure minimizes the squared error between the measured received power and the model's predicted values, allowing us to calibrate the model parameters based on real-world data.

To validate the effectiveness of our altitude-dependent model, we compare it against the classical FSPL model defined in~\eqref{eq:PL_fs}. Fig.~\ref{fig:cdf_err_packa} shows the empirical cumulative distribution function (CDF) of modeling errors for both models across urban and rural scenarios. Our model demonstrates significantly lower error in urban environments, particularly at low altitudes, where NLoS propagation effects are dominant. This confirms the limitations of FSPL in complex urban topologies and underscores the importance of incorporating altitude-aware corrections.

\begin{table*}[t!]
\caption{Received signal power analysis by using the FSPL model in Lake Wheeler (rural) and Packapaloza (urban) datasets.}
\label{table:fre_spac}
\centering
\scalebox{0.95}{
\begin{tabular}{p{0.8cm}|p{1.5cm}|p{1.4cm}|p{1.7cm}|p{2.2cm}|p{2.4cm}|p{2.2cm}|p{2.4cm}}
\hline
\textbf{Call Sign}& Experiment site & Free space path loss (dB)& Analytical received signal power (dBm)& Measured received signal power (dBm) in 2023& Offset between analysis and measurement (dB) in 2023& Measured received signal power (dBm) in 2022& Offset between analysis and measurement (dB) in 2022\\
\hline\hline
\textbf{WKNC}& \centering Packapalooza& \centering $-62.42$& \centering $11.56$& \centering $40.93$& \centering $29.37$& \centering N/A & N/A \\
\hline
\textbf{WNCB}& \centering Packapalooza& \centering $-95.88$& \centering $-15.88$& \centering $18.78$& \centering $34.66$& \centering $15.83$& $31.71$\\
\hline
\textbf{WQDR}& \centering Packapalooza& \centering $-96.59$& \centering $-16.81$& \centering $17.85$& \centering $34.66$& \centering $15.68$&  $32.49$\\
\hline
\textbf{WBBB}& \centering Packapalooza& \centering $-93.85$& \centering $-13.94$& \centering $20.43$& \centering$34.37$& \centering $24.00$&  $37.94$\\
\hline\hline
\textbf{WKNC}& \centering LakeWheeler& \centering $-88.39$& \centering $-14.41$& \centering N/A & \centering N/A &\centering $11.59$&  $26.00$\\
\hline
\textbf{WNCB} & \centering LakeWheeler& \centering $-92.76$& \centering $-12.76$& \centering N/A & \centering N/A &\centering $16.44$&  $29.20$\\
\hline
\textbf{WQDR} & \centering LakeWheeler& \centering $-95.83$& \centering $-16.05$& \centering N/A & \centering N/A &\centering $14.10$&  $30.15$\\
\hline
\textbf{WBBB} & \centering LakeWheeler& \centering $-86.31$& \centering $-6.40$& \centering N/A & \centering N/A &\centering $26.77$&  $33.17$\\
\hline\hline
\end{tabular}
}
\end{table*}

\subsection{Path Loss Analysis on Rural Measurement Datasets}
Fig.~\ref{fig:los_alt_LW} shows the received signal power from different FM radio stations obtained from the 2022 Lake Wheeler dataset~\cite{maeng2023spectrum}. As illustrated in Fig.~\ref{fig:wknc_cov}, the Lake Wheeler site is a rural area with minimal urban obstruction. Consequently, signals from stations located to the southwest and south (e.g., WNCB and WBBB) exhibit strong and consistent received power even from low altitudes, indicating LoS conditions starting as low as $10$~m. By contrast, the signal from WQDR (southeast direction) gradually increases until about $80$~m, indicating a delayed LoS transition due to semi-urban obstructions. WKNC, located in a more urban northern direction with a relatively low antenna height, exhibits a gradual increase until approximately $120$~m.

In Fig.~\ref{fig:los_alt_LW_ana}, we apply the FSPL model to WNCB and WBBB (consistent LoS) and the break-point path loss model to WKNC and WQDR (height-dependent LoS). Fitting parameters $\alpha = 4.5$ for WKNC and $6$ for WQDR yield the best fit. A $4$~dB compensation offset is applied to WKNC to account for its antenna radiation pattern as explained in Table~\ref{table:fre_spac}. The CDF of modeling error in Fig.~\ref{fig:cdf_err_lw} confirms these model selections—WKNC and WQDR show improved performance with the break-point model, while WNCB and WBBB align better with FSPL.

These results highlight a critical insight: altitude plays a key role in signal propagation dynamics, even for low-frequency FM transmissions. In urban areas, the transition from NLoS to LoS occurs at a higher altitude threshold due to obstructions, which is effectively captured by our proposed model. In rural settings, LoS is typically present from lower altitudes, and FSPL suffices in many cases. These findings emphasize the importance of environment-specific, altitude-aware path loss modeling in spectrum prediction tasks.

\subsection{Spectrum Analysis Based on the LoS and NLoS Conditions}
Based on the LoS altitude threshold ($50$~m) identified earlier, we analyze the spectrum activity of FM radio band considering LoS/NLoS conditions. Fig.~\ref{fig:spec_los_nlos}  shows the spectrum activity of LoS and NLoS conditions in terms of various statistics including mean, standard deviation, median, and CDF. In Fig.~\ref{fig:spec_mean_los}, we clearly see four peaks of signal power, which are indicated by corresponding FM radio stations. In addition, the standard deviation is relatively small compared with peak mean values.

In Fig.~\ref{fig:spec_cdf_los}, all the power values from the measurements are marked and the color of the marked measurement points is mapped to the CDF value of the power at the specific frequency. Note that the median curves represent the power at CDF value being $0.5$. It is observed that the distribution of signal power is relatively uniform in the LoS condition.

In Fig.~\ref{fig:spec_mean_nlos}, we observe that the mean of the NLoS condition is lower than that of the LoS condition shown in Fig.~\ref{fig:spec_mean_los}, while the standard deviation of the NLoS is higher than that of the LoS condition. It implies that the blockage in the NLoS condition reduces the signal power strength and scatters the reflected signals.

In Fig.~\ref{fig:spec_cdf_nlos}, It is observed that the distribution of signal power is relatively broad in the NLoS condition compared with the LoS condition shown in Fig.~\ref{fig:spec_cdf_los}. Due to the reflection and scattering, a wide range of signal power distribution is observed in the NLoS condition. Moreover, we observe the abruptly increasing maximum power between $88.5$~MHz and $94$~MHz. This is because the location of the helikite is continuously changed by the wind, and the NLoS condition from obstacles can be suddenly released to signals from FM radio stations in this band. In addition, we would capture a more gradual increase in the signal power, if the sweep duration is shorter than around 15 seconds in this experiment.

\subsection{Free Space Path Loss Analysis}\label{sec:fre_spa}

In the LoS link condition, the received signal strength can be modeled by the free space path. Moreover, since both the Tx antenna and the Rx antenna are placed above the building height at the link between FM radio stations and the heikite, it is fairly reasonable to adopt free space path loss analysis in \eqref{eq:PL_fs}. By using information of FM radio stations in Table~\ref{table:FM_radi_sta} and the mean of measured signal power in the LoS condition in Fig.~\ref{fig:spec_mean_los}, we compare the received signals from the analysis and the measurement summarized in Table~\ref{table:fre_spac}. It is worth noting that we utilize Packapalooza and Lake Wheeler helikite measurement datasets from 2022 as well as the packpalooza dataset from 2023 in analyzing the received signal strength by free space path loss, which can be found in~\cite{AERPAW}. 

In the 2023 Packapalooza dataset, we observe that the offsets between the received signal strength from the measurement and the free space path loss analysis are around $29$~dB in WKNC, while around $34$~dB in WNCB, WQDR, and WBBB. Since similar offset values are observed from different FM radio stations, we can conclude that these offsets dominantly come from the calibration error in the SDR receiver.  However, the offset of WKNC is around $4$~dB less than the other three FM radio stations. We may interpret the reason as the antenna gain not being maximized in the direction of the helikite due to the directional antenna pattern shown in Fig.~\ref{fig:wknc_cov}. In the 2022 Packapalooza dataset, the offset is around $32$~dB in WNCB and WQDR while around $38$~dB in WBBB, which shows that the calibration error of the SDR receiver would be varied for each experiment. In the 2022 Lake Wheeler dataset, the offset is around $30$~dB in WNCB and WQDR while around $33$~dB in WBBB. In future work, we plan to measure the calibration error of the SDR receiver using dedicated calibration equipment prior to data collection.

\section{Conclusion}\label{sec:conclusion}
In this paper, we present spectrum monitoring experiment results using a helikite-mounted SDR in an urban area. Based on information on the location and effective radiated power near FM radio stations, we analyze the received signal strength of the FM radio signals captured by the spectrum monitoring. We show the altitude-dependent spectrum power and CDF of the measured signal power across the spectrum. In addition, we analyze LoS and NLoS conditions depending on the altitude, and propose an altitude-dependent path loss model and fit it to the measurement results. We calibrate the offset of the received signal power from the SDR receiver by using the free space path loss analysis of FM radio signals. Ultimately, this study establishes a scalable modeling framework and validation methodology that can be extended to characterize the 3D propagation dynamics of emerging aerial technologies, such as 5G and 6G networks, across diverse frequency bands.

\section{Acknowledgment}
The authors would like to thank Thomas Zajkowski and Evan Arnold for their help with the helikite experiments. 
\bibliographystyle{IEEEtran}
\bibliography{IEEEabrv,references}

\begin{thebibliography}{10}
\providecommand{\url}[1]{#1}
\csname url@samestyle\endcsname
\providecommand{\newblock}{\relax}
\providecommand{\bibinfo}[2]{#2}
\providecommand{\BIBentrySTDinterwordspacing}{\spaceskip=0pt\relax}
\providecommand{\BIBentryALTinterwordstretchfactor}{4}
\providecommand{\BIBentryALTinterwordspacing}{\spaceskip=\fontdimen2\font plus
\BIBentryALTinterwordstretchfactor\fontdimen3\font minus \fontdimen4\font\relax}
\providecommand{\BIBforeignlanguage}[2]{{%
\expandafter\ifx\csname l@#1\endcsname\relax
\typeout{** WARNING: IEEEtran.bst: No hyphenation pattern has been}%
\typeout{** loaded for the language `#1'. Using the pattern for}%
\typeout{** the default language instead.}%
\else
\language=\csname l@#1\endcsname
\fi
#2}}
\providecommand{\BIBdecl}{\relax}
\BIBdecl

\bibitem{maeng2022national}
S.~J. Maeng, I.~G{\"u}ven{\c{c}}, M.~Sichitiu, B.~A. Floyd, R.~Dutta, T.~Zajkowski, {\"O}.~{\"O}zdemir, and M.~J. Mushi, ``National radio dynamic zone concept with autonomous aerial and ground spectrum sensors,'' in \emph{Proc. IEEE Int. Conf. Commun. (ICC) Workshops}, Seoul, Korea, May 2022.

\bibitem{8254105}
A.~Selim, F.~Paisana, J.~A. Arokkiam, Y.~Zhang, L.~Doyle, and L.~A. DaSilva, ``Spectrum monitoring for radar bands using deep convolutional neural networks,'' in \emph{Proc. IEEE Global Telecommun. Conf.}, Singapore, Dec. 2017, pp. 1--6.

\bibitem{9395503}
F.~A. Bhatti, M.~J. Khan, A.~Selim, and F.~Paisana, ``Shared spectrum monitoring using deep learning,'' \emph{IEEE Trans. Cogn. Commun. Netw.}, vol.~7, no.~4, pp. 1171--1185, Dec. 2021.

\bibitem{7343894}
M.~Souryal, M.~Ranganathan, J.~Mink, and N.~E. Ouni, ``{Real-time centralized spectrum monitoring: Feasibility, architecture, and latency},'' in \emph{Proc. IEEE Int. Symp. Dynam. Spectrum Access Netw.}, Stockholm, Sweden, Sep. 2015, pp. 106--112.

\bibitem{raouf2023cellular}
A.~H.~F. Raouf, S.~J. Maeng, I.~Guvenc, {\"O}.~{\"O}zdemir, and M.~Sichitiu, ``{Cellular Spectrum Occupancy Probability in Urban and Rural Scenarios at Various UAS Altitudes},'' in \emph{Proc. IEEE Int. Symp. Pers., Indoor, Mobile Radio Commun.}, Toronto, Canada, Sep. 2023.

\bibitem{maeng2023sdr}
S.~J. Maeng, O.~Ozdemir, {\.I}.~G{\"u}ven{\c{c}}, M.~L. Sichitiu, M.~Mushi, R.~Dutta, and M.~Ghosh, ``{SDR-Based 5G NR C-Band I/Q Monitoring and Surveillance in Urban Area Using a Helikite},'' in \emph{Proc. IEEE Int. Conf. Ind. Technol.}, Orlando, FL, USA, Apr. 2023, pp. 1--6.

\bibitem{maeng2023spectrum}
S.~Maeng, O.~Ozdemir, H.~Nandakumar, {\.I}.~G{\"u}ven{\c{c}}, M.~Sichitiu, R.~Dutta, and M.~Mushi, ``{Spectrum activity monitoring and analysis for sub-6 GHz bands using a Helikite},'' in \emph{Proc. Int. Conf. Commun. Syst. Netw.}, Bangalore, India, Jan. 2023, pp. 857--862.

\bibitem{CellMapper}
\BIBentryALTinterwordspacing
``{CellMapper: Cellular coverage and tower map}.'' [Online]. Available: \url{https://www.cellmapper.net/}
\BIBentrySTDinterwordspacing

\bibitem{radio_locator}
\BIBentryALTinterwordspacing
``{Radio-locator in Raleigh NC USA}.'' [Online]. Available: \url{https://radio-locator.com/cgi-bin/locate?select=city&city=Raleigh&state=NC}
\BIBentrySTDinterwordspacing

\bibitem{FCC_public_file}
\BIBentryALTinterwordspacing
``{FCC public inspection file of FM station WKNC}.'' [Online]. Available: \url{https://publicfiles.fcc.gov/fm-profile/WKNC-FM/contour-maps}
\BIBentrySTDinterwordspacing

\bibitem{AERPAW}
\BIBentryALTinterwordspacing
``{APERPAW experiments dataset}.'' [Online]. Available: \url{https://aerpaw.org/experiments/datasets/}
\BIBentrySTDinterwordspacing

\end{thebibliography}
  
\end{document}